\documentclass[%
 reprint,
unsortedaddress,
 amsmath,amssymb,
 aps,
]{revtex4-2}
\usepackage{graphicx}
\usepackage{dcolumn}
\usepackage{bm}
\usepackage{xcolor}

\newcommand{\ket}[1]{\left|#1\right\rangle}

\newcommand{\braket}[3]{\left\langle #1\left| #2 \right| #3\right\rangle}

\usepackage{soul}

\begin{document}

\title{Degeneracy in excited-state quantum phase transitions of two-level bosonic models and its influence on system dynamics}

\author{Jamil Khalouf-Rivera}
\affiliation{Departamento de Ciencias Integradas y Centro de Estudios Avanzados en Física, Matemáticas y Computación, Universidad de Huelva, Huelva 21071, Spain}
\affiliation{School of Physics, Trinity College Dublin, College Green, Dublin 2, Ireland}

\author{Qian Wang}
\affiliation{Department of Physics, Zhejiang Normal University, Jinhua 321004, China}
\altaffiliation[Also at ]{CAMTP-Center for Applied Mathematics and Theoretical Physics, University of Maribor, 
Mladinska 3, SI-2000 Maribor, Slovenia}

\author{Lea F.~Santos}
\affiliation{Department of Physics, University of Connecticut, Storrs, Connecticut, USA}

\author{José-Enrique~García-Ramos}
\affiliation{Departamento de Ciencias Integradas y Centro de Estudios Avanzados en Física, Matemáticas y Computación, Unidad Asociada GIFMAN, CSIC-UHU, Universidad de Huelva, Huelva 21071, Spain}
\altaffiliation[Also at ]{Instituto Carlos I de F\'{\i}sica Te\'orica y Computacional, Universidad de Granada, Fuentenueva s/n, 18071 Granada, Spain}

\author{Miguel~Carvajal}
\affiliation{Departamento de Ciencias Integradas y Centro de Estudios Avanzados en Física, Matemáticas y Computación, Unidad Asociada GIFMAN, CSIC-UHU, Universidad de Huelva, Huelva 21071, Spain}
\altaffiliation[Also at ]{Instituto Carlos I de F\'{\i}sica Te\'orica y Computacional, Universidad de Granada, Fuentenueva s/n, 18071 Granada, Spain}

\author{Francisco~Pérez-Bernal}
\affiliation{Departamento de Ciencias Integradas y Centro de Estudios Avanzados en Física, Matemáticas y Computación, Unidad Asociada GIFMAN, CSIC-UHU, Universidad de Huelva, Huelva 21071, Spain}
\altaffiliation[Also at ]{Instituto Carlos I de F\'{\i}sica Te\'orica y Computacional, Universidad de Granada, Fuentenueva s/n, 18071 Granada, Spain}


\date{\today}

\begin{abstract}
Excited-state quantum phase transitions (ESQPTs) strongly influence the spectral properties of collective many-body quantum systems, changing degeneracy patterns in different quantum phases. Level degeneracies, in turn, affect the system's dynamics. We analyze the degeneracy dependence on the size of two-level boson models with a $u(n+1)$ dynamical algebra, where $n$ is the number of collective degrees of freedom. Below the ESQPT critical energy of these models,  the energy gap between neighboring levels that belong to different symmetry sectors  gets close to zero as the system size increases. We report and explain why this gap goes to zero exponentially for systems with one collective degree of freedom, but algebraically in models with more than one degree of freedom. As a consequence, we show that the infinite-time average of out-of-time-order correlators is an ESQPT order parameter in finite systems with $n=1$, but in systems with $n>1$, this average only works as an order parameter in the mean-field limit. 
\end{abstract}

\maketitle


\section{Introduction}
\label{sec_intro}

Quantum phase transitions (QPTs) are zero-temperature phase transitions driven by quantum instead of thermal fluctuations. In QPTs the system's ground state~\footnote{This is why such transitions are also known as ground-state quantum phase transitions.} undergoes an abrupt change once one or several Hamiltonian parameters -- denoted as control parameters --  reach particular critical values~\cite{Carr2010}. In collective quantum systems, true discontinuities only happen in the large system size limit (also known as the mean-field limit). However, QPT precursors can be found in finite-size systems~\cite{Iachello2004}. The QPT concept was later extended to encompass excited states, with the introduction of what became known as excited-state quantum phase transitions (ESQPTs), which are marked by discontinuities in the excited states energy level density and in its dynamics as a function of the control parameter (excited energy level flow)  at critical values of the energy~\cite{Cejnar2006, Caprio2008, Cejnar2008}. An extensive review paper on ESQPTs has been recently published, where the interested reader can find the main developments in the field \cite{[][{, and references therein.}]Cejnar2021}. 

In the present work, we deal with two-level boson models with a $u(n+1)$ Lie algebra as their dynamical algebra, where $n$ is the number of the system's collective degrees of freedom. The dynamical algebra generators are built as bilinear products of creation and annihilation operators of a scalar boson plus an $n$-dimensional boson operator~\cite{general0}. Within the algebraic approach, the possible chains of subalgebras that start in the \(u(n+1)\) dynamical algebra and end in the system's symmetry algebra 
are called \textit{dynamical symmetries}. They provide analytically solvable examples of physical limits of interest, as well as bases to carry out calculations~\cite{bookalg,frank}. These models make extensive use of symmetries and provide good approximations to the collective degrees of freedom of a variety of complex quantum systems. They have been successfully applied to the collective degrees of freedom of nuclei (interacting boson model (IBM)~\cite{booknuc, ibfm_book}), molecules (vibron model~\cite{bookmol}), and baryons (algebraic approach to baryons~\cite{BIJKER1994, BIJKER2000}). QPTs~\cite{Cejnar2010} and ESQPTs~\cite{Cejnar2021} have attracted a great deal of attention in algebraic models since the pioneering works that introduced the concept of shape-phase transitions in nuclei~\cite{Gilmore1979, Feng1981}. The ground-state QPTs for collective quantum systems were classified in Ref.~\cite{Cejnar2007}, considering a general system Hamiltonian including one- and two-body interactions.

In the main text, we analyze a two-level boson model with $n=1$, the Lipkin-Meshkov-Glick (LMG) model, and a model with $n=2$, the two-dimensional limit of the vibron model (2DVM). In Appendix \ref{appa}, we extend our results to $n = 3$ and to $n=5$ models [the vibron model (VM) and the interacting boson model (IBM)]. 
All of these models exhibit a second-order ground-state QPT between their \(u(n)\) and \(so(n+1)\) dynamical symmetries; their mean-field limit  can be explored  using the coherent-state formalism and the resulting energy functional indicates that their QPTs can be explained using a cusp catastrophe model~\cite{Cejnar2007}.

The LMG model and the 2DVM (as well as the models presented in the Appendix \ref{appa})  display an ESQPT in the broken-symmetry phase of the second-order ground-state QPT.  The ESQPT precursors of these collective quantum models display many similarities, such as the abrupt localization at the ESQPT energy of the systems' eigenstates expressed in the basis associated with the \(u(n+1)\supset u(n)\) dynamical symmetry~\cite{Santos2016, PBernal2017,Santos2015}. It would then seem plausible to expect that ESQPT precursors and their influence on system dynamics would be common to the different models under consideration. Yet, we show that this is not the case. In  the broken-symmetry phase below the ESQPT, pairs of excited levels, each level belonging to a different symmetry sector, are (nearly) degenerate for both models, but the nature of this degeneracy is qualitatively different~\cite{Caprio2008}.

In models with \(n=1\), the energy splitting between two energy levels decreases exponentially as the system size increases, while for models with \(n>1\), the splitting decreases algebraically with the system size.
This difference should greatly affect the broken-symmetry phase dynamics. We illustrate this point with the calculation of the long-time averaged value of a microcanonical out-of-time order correlator (mOTOC)~\cite{Hashimoto2017} for the LMG model and the 2DVM. We confirm that this quantity, for systems of finite size, can be used as an order parameter for the ESQPT in the LMG model, as proposed in~\cite{Wang2019b, KRivera2022b}, but we show that it is only an approximate order parameter for the 2DVM and other models with $n>1$.

The present work is structured as follows. Section~\ref{sec2}  describes the LMG and the 2DVM model; Sec.~\ref{sec3} presents our main results for the degeneracy of states in the broken-symmetry phase of both models; and in Sec.~\ref{sec4}, through the calculation of the long-time averaged value of a mOTOC, we show how the different nature of the degeneracies associated with each model  affects their dynamics. Our conclusions are in Sec.~\ref{Sec:Conclusions}.

\section{Selected two-level boson models}
\label{sec2}

As mentioned above, we are dealing with two-level boson models with a \(u(n+1)\) Lie algebra as their dynamical algebra. Such model have as a basic ingredient two types of boson operators. The first boson operator is a scalar boson and the second one is an \(n\)-dimensional tensor operator. The number \(n\) is determined by the number of collective degrees of freedom of the system under study. The \((n+1)^2\) generators of the dynamical algebra are built as the possible products of a creation and an annihilation boson operator~\cite{bookalg}. In the main text, we present results for the LMG and the 2DVM models. In the LMG case, the system has a single collective degree of freedom, \(n = 1)\), hence both bosons are scalar and the  dynamical algebra is \(u(2)\)\footnote{The  second boson is considered a pseudoscalar one as discussed in~\cite{frank}.}. The 2DVM was introduced for the study of systems with two degrees of freedom, \(n=2\) and the scalar boson is combined with two Cartesian bosons. The models introduced in the Appendix~\ref{appa} are the VM and the IBM. The first one was introduced to model the rovibrational spectrum of diatomic molecules, where the main ingredient is the dipole, and hence \(n=3\). Therefore, the scalar boson is combined with the three components of a vector boson and the dynamical algebra is \(u(4)\). In the IBM case, devised to model collective surface vibrations of nuclei, the scalar boson is combined with a quadrupolar boson (angular momentum two) and the resulting dynamical algebra is \(u(6)\). We proceed to briefly describe the LMG and the 2DVM.

\subsection{The Lipkin-Meshkov-Glick model}
\label{Sec:LMG}
The LMG model has a single effective degree of freedom ($n = 1$). It was introduced in the 60's as a toy model for the study of the validity of approximate methods in nuclear structure studies~\cite{Lipkin1965,Meshkov1965,Glick1965}. The LMG model has been extensively used in the study of  QPTs~\cite{Dusuel2004,Castanos2006,Ribeiro2008,Engelhardt2013,Romera2014,Heyl2018} and ESQPTs~\cite{Heiss2005, Relano2008, PFernandez2009, PFernandez2011, Yuan2012, Santos2016, Wang2019a, Wang2019b, Kelly2020, Wang2021, Nader2021, Gamito2022, KRivera2022b}. In addition to its simplicity and rich physical content, this model can be realized with a fully-connected chain of $N$ spins~\cite{Santos2016} and it has several experimental realizations~\cite{Zibold2010, AFerreira2013, Jurcevic2014, Jurcevic2017, Muniz2020, Makhalov2019,Cervia2021,Hlatshwayo2022,Li2023}.

The general LMG model has first-, second-, and third-order ground-state QPTs \cite{Castanos2006, Romera2014}. Here, we work with the following simplified LMG Hamiltonian
\begin{equation}
  \hat H_{LMG} = (1-\xi)\left(\frac{N}{2} + \hat J_z\right) - \frac{4\xi}{N}  \hat J_x^2~,
  \label{HLMGM}
\end{equation}
\noindent 
where $J_z$, $J_x$ are quasispin components, and \(\xi\) is  a control parameter, \(\xi\in[0,1]\). This Hamiltonian is written in an intensive form, dividing the second term (a two-body interaction), by the system size, $N$, to facilitate the access to the large size limit of the system.

The algebraic structure of this model is made clear in its bosonic realization, introducing scalar, \(s\), and pseudoscalar, \(t\), bosons. The four generators of the $u(n+1) = u(2)$ dynamical algebra are the bilinear products \(\{s^\dagger s, t^\dagger t, t^\dagger s, s^\dagger t\}\) and, making use of the Schwinger representation, the generators can be recast as the quasispin components \(\{\hat N, \hat J_x, \hat J_y, \hat J_z\}\) \cite{bookalg, frank} 
\begin{align*}
  \hat N &= t^\dagger t + s^\dagger s~, &\hat J_x &= \frac{1}{2}(t^\dagger s + s^\dagger t)~,\\
  \hat J_y &=  \frac{1}{2i}(t^\dagger s - s^\dagger t) ~, &\hat J_z &=  \frac{1}{2}(t^\dagger t - s^\dagger s)~.
\end{align*}
\noindent The \(\hat N\) operator in this case is constant and equal to the total number of \(s\) and \(t\) bosons, denoting the totally symmetric \(u(2)\) irrep that spans the system's Hilbert space. There are two possible dynamical symmetries starting from the $u(2)$ dynamical algebra
\begin{align}
&u(2)\supset u(1)& \mbox{Chain (LMG-I)}~,\label{lmgmi}\\
&u(2)\supset so(2) &\mbox{Chain (LMG-II)}~.\label{lmgmii}
\end{align}
The Hamiltonian Eq.~\eqref{HLMGM} can then be recast as 
\begin{equation}
  \hat H_{LMG} = (1-\xi)\hat n_t + \frac{\xi}{N}\hat P_t~,
  \label{HLMGM_boson}
\end{equation}
\noindent where $\hat n_t = t^\dagger t$ and $\hat P_t = N^2 - (t^\dagger s + s^\dagger t)^2$.  Hence, the model Hamiltonian combines the first-order Casimir operator of $u(1)$ and the second-order Casimir operator of $so(2)$. The LMG-I(II) dynamical symmetry is recovered for \(\xi = 0(1)\). 

The simplified \(\hat H_{LMG}\) in Eq.~\eqref{HLMGM} can be split into even and odd symmetry blocks, as it conserves parity \(\hat \Pi = e^{\imath \pi \hat{n}_t }\). Hamiltonian (\ref{HLMGM}) has a second-order ground-state QPT for a critical value of the control parameter, \(\xi_c = 0.2\), and an associated ESQPT  \cite{Heiss2005, Relano2008, Santos2016}.

\subsection{The two-dimensional limit of the vibron model}
\label{Sec:2DVM}

The 2DVM was first introduced for the study of vibrational bending degrees of freedom in molecules \cite{Iachello1996}. Molecular bending is a planar motion and it implies two degrees of freedom, hence \(n = 2\) and the system dynamical algebra is the $u(3)$ Lie algebra. In this approach, bending vibrations are treated as collective bosonic excitations and the model building blocks are a scalar boson operator,  \(\sigma\), and two circular bosons, \(\tau_i\) with \(i = +,-\) \cite{PBernal2008}. As in the LMG model case, the nine  \(u(3)\) generators are the possible bilinear products of creation and annihilation operators \cite{Iachello1996,PBernal2008}. The 2DVM is the simplest two-level model with nontrivial angular momentum, a fact that has made it a convenient model for QPT and ESQPT studies \cite{PBernal2005, Caprio2008, PBernal2008, a2DVM_2022, Novotny2023, Corps2024}. It is worth to emphasize that the first experimental signatures of ESQPT precursors were found in the molecular bending spectrum of nonrigid molecules \cite{Larese2011,Larese2013,KRivera2020}.

As in the LMG model case, the 2DVM has two dynamical symmetries. In this case, considering angular momentum conservation, both dynamical symmetries converge in the system's symmetry algebra, $so(2)=\left\{\hat{\ell}=\tau_+^{\dagger}\tau_+-\tau_-^{\dagger}\tau_-\right\}$,
\begin{align}
&u(3)\supset u(2) \supset so(2)&\mbox{Chain (2DVM-I)}~,\label{2dvmi}\\
&u(3)\supset so(3)\supset so(2) &\mbox{Chain (2DVM-II)}~.\label{2dvmii}
\end{align}

In this model, the total number of \(\sigma\) and \(\tau\) bosons is denoted as \(N\), a constant that determines the totally symmetric irrep of \(u(3)\) that spans the system's Hilbert space. The dynamical symmetry in Eq.~\eqref{2dvmi} is a convenient approximation to model the bending degrees of freedom of linear molecules. The second dynamical symmetry, Eq.~\eqref{2dvmii}, is applied to the modeling of rigidly-bent molecular species \cite{Iachello1996, PBernal2008, KRivera2020}.  

In the same spirit as we have done for the LMG model, we introduce a simple model Hamiltonian that allows for the transition between the limiting cases associated with the two dynamical symmetries of the 2DVM,
\begin{equation}
  \hat H_{2DVM} = (1-\xi) \hat n_\tau + \frac{\xi}{N}  \hat P_\tau~,
  \label{H2DVM}
\end{equation}
\noindent where the control parameter \(\xi\in[0,1]\) and the operator \(\hat n_\tau = \tau^\dagger_+\tau_+ + \tau^\dagger_-\tau_-\) is the first-order Casimir operator of the \(u(2)\) subalgebra in Eq.~\eqref{2dvmi}. The pairing operator \(\hat P_\tau =  N(N+1) - \hat W^2\), where \(\hat W^2=\frac{1}{2}\left(\hat{D}_+\hat{D}_-+\hat{D}_-\hat{D}_+\right)+\hat{\ell}^2\) is the second-order Casimir operator of the subalgebra \(so(3)=\text{span}\left\{\hat{D}_{\pm}=\sqrt{2}\left(\pm \tau_\pm^{\dagger}\sigma \mp \sigma^{\dagger}\tau_{\mp}\right),\hat{\ell}\right\}\) in Eq.~\eqref{2dvmii}~\cite{PBernal2008}. 

Considering the conservation of the system angular momentum, \(\ell\) \footnote{We used the notation $\ell$ for the \(so(2)\) quantum number, that is commonly used in molecular spectroscopy for the vibrational angular momentum in the doubly-degenerate bending degree of freedom of linear species.}, the model Hamiltonian is block-diagonal for states belonging to different irreps of the symmetry algebra \(so(2)\).
For molecular bending vibrations, this conserved quantity 
can be identified either with the vibrational angular momentum --in the \(u(2)\) dynamical symmetry-- or with the projection of the angular momentum on the molecular figure axis --in the \(so(3)\) dynamical symmetry \cite{PBernal2008, KRivera2020}. 

\section{Results}
\label{sec3}
\begin{figure*}
    \includegraphics[width=0.75\textwidth, angle=0]{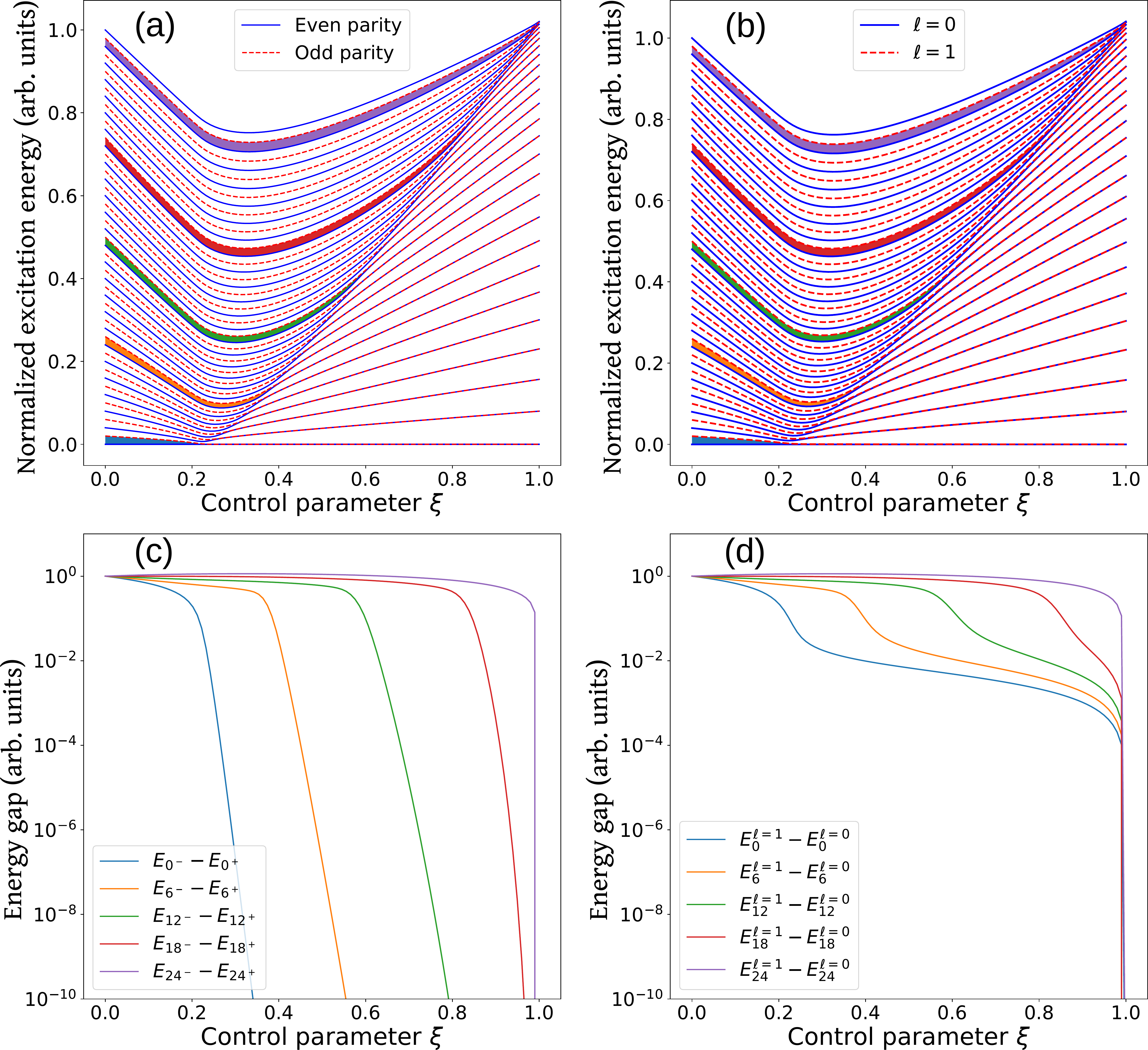}
\caption{\label{lmg_2dvm_ced}  Panel (a): Excitation energy scaled by the system size ($\varepsilon = (E-E_0)/N$) as a function of the control parameter $\xi$ for the LMG model with model Hamiltonian Eq.~\eqref{HLMGM_boson} and system size \(N = 50\). Solid blue (dashed red) lines mark even (odd) parity levels.  Panel (b): Excitation energy scaled by the system size ($\varepsilon = (E-E_0)/N$) as a function of the control parameter $\xi$ for the 2DVM model Hamiltonian Eq.~\eqref{H2DVM} with a system size \(N = 50\). Solid blue (dashed red) lines mark levels with angular momentum \(\ell = 0 (1)\). In both panels the color-filled area marks the energy difference  between selected states with different parity (panel (a)) or angular momentum (panel (b)). Panel (c): Energy difference between selected pairs of states of the LMG model Hamiltonian Eq.~\eqref{HLMGM_boson} having different parity as a function of the control parameter $\xi$. Panel (d): Energy difference between selected states of Hamiltonian Eq.~\eqref{H2DVM} with angular momentum \(\ell = 0\) and \(1\) as a function of the control parameter $\xi$. In both cases, the labels of the selected pairs of levels are provided in the legend of the panels and we use for each pair of states the same color used to fill the corresponding area in the upper panels.}
\end{figure*}

 The ESQPT in the bosonic models considered here is characterized by the divergence of the density of states at the ESQPT critical energy. This is illustrated for the LMG and the 2DVM in  Figs.~\ref{lmg_2dvm_ced}(a) and~\ref{lmg_2dvm_ced}(b). The excitation energy, scaled by the system size, is depicted as a function of the $\xi$ control parameter for the LMG and 2DVM  with the same system size \(N = 50\). For each value of $\xi$ larger than the critical value (\(\xi_c = 0.2\)), the point where the energy levels accumulate marks the critical energy of the ESQPT and the line of maximum density of states is the separatrix between the ESQPT phases.

At energies below the critical energy, it
 can be clearly appreciated  in  Fig.~\ref{lmg_2dvm_ced}(a) how even and odd parity states (depicted with solid blue and dashed red lines, respectively) are degenerate in the broken-symmetry region ($\xi_c < \xi < 1$) at energies less than the critical energy. This degeneracy is broken for states with energies greater than the critical ESQPT energy. This phenomenon, also dubbed as \textit{level kissing}, has recently been experimentally accessed using a squeeze-driven Kerr oscillator realized with a superconducting circuit \cite{Frattiniarxiv} and the corresponding ESQPT features have been identified in~\cite{Chavez2023}. 
 
 The level degeneracy at energies below the critical energy is common to ESQPTs in different systems and it has been used to define a constant of the motion able to identify ESQPT dynamic phases in quantum collective models with a single degree of freedom \cite{Corps2021,Corps2022}. The 2DVM correlation energy diagram is shown in Fig.~\ref{lmg_2dvm_ced}(b) with a clear similarity to Fig.~\ref{lmg_2dvm_ced}(a). In this case, excitation energies  for states with angular momentum \(\ell = 0\) (solid blue lines) and \(1\) (dashed red lines) are plotted as a function of the control parameter, \(\xi\), for a system size \(N = 50\).  At first sight, the results for both models seem to be completely equivalent, something that should not be surprising considering that both cases have a second-order ground-state QPT at $\xi_c$ and an ESQPT associated with the ground-state transition.
 
However, as already noted by Caprio and collaborators in Ref.~\cite{Caprio2008}, the energy gap between even- and odd-parity state pairs in the LMG model is much smaller than the energy difference between the corresponding states with different angular momentum values in the 2DVM case for a common system size. To illustrate this point, we select for each model four pairs of states with different symmetry  located at different excitation energies in Figs.~\ref{lmg_2dvm_ced}(a) and \ref{lmg_2dvm_ced}(b). We  highlight the energy difference between the states with blue, orange, green, red, and purple colors. The blue color is used to fill the energy gap that exists between the system ground state and the first odd state in the LMG case and between the ground state and the first \(\ell = 1\) state in the 2DVM case. As  the control parameter increases from zero and the system gets closer to the critical ESQPT energy, the width of the colored surface decreases,  disappearing once the pair of levels cross the ESQPT separatrix.

The difference between the two models is clearly evinced once the energy gap between the selected level pairs (note that this  quantity is not scaled by the system size) is depicted, using logarithmic scale and the same colors, in Figs.~\ref{lmg_2dvm_ced}(c) and \ref{lmg_2dvm_ced}(d). 
In the LMG case, shown in  Fig.~\ref{lmg_2dvm_ced}(c), the energy difference tends to zero once the separatrix is crossed and soon it becomes less than the numerical precision used in the calculations, even for a finite-size system. However, in the 2DVM, shown in Fig.~\ref{lmg_2dvm_ced}(d), the energy difference between adjacent states with \(\ell = 0\) and \(\ell = 1\) experiences a fast decrease as the separatrix is crossed, but then it smoothly decreases to values between \(10^{-3}\) and \(10^{-2}\). For this model, the levels become exactly degenerate only when the control parameter is $\xi=1$, right in the $so(3)$ dynamical symmetry, where solutions are analytic and different angular momenta values are known to collapse in the vibrational head \cite{PBernal2008}. The comparison of the results in panels Figs.~\ref{lmg_2dvm_ced}(c) and \ref{lmg_2dvm_ced}(d) reveals a fundamental difference between the model with \(n=1\) and the model with \(n=2\). Energy differences for adjacent states with different symmetry tend to zero in the LMG model, even for finite-size systems, while in the 2DVM case the corresponding energy differences are expected to decrease with the system size, becoming zero only in the large system size limit.

\begin{figure*}
\includegraphics[width=0.75\textwidth]{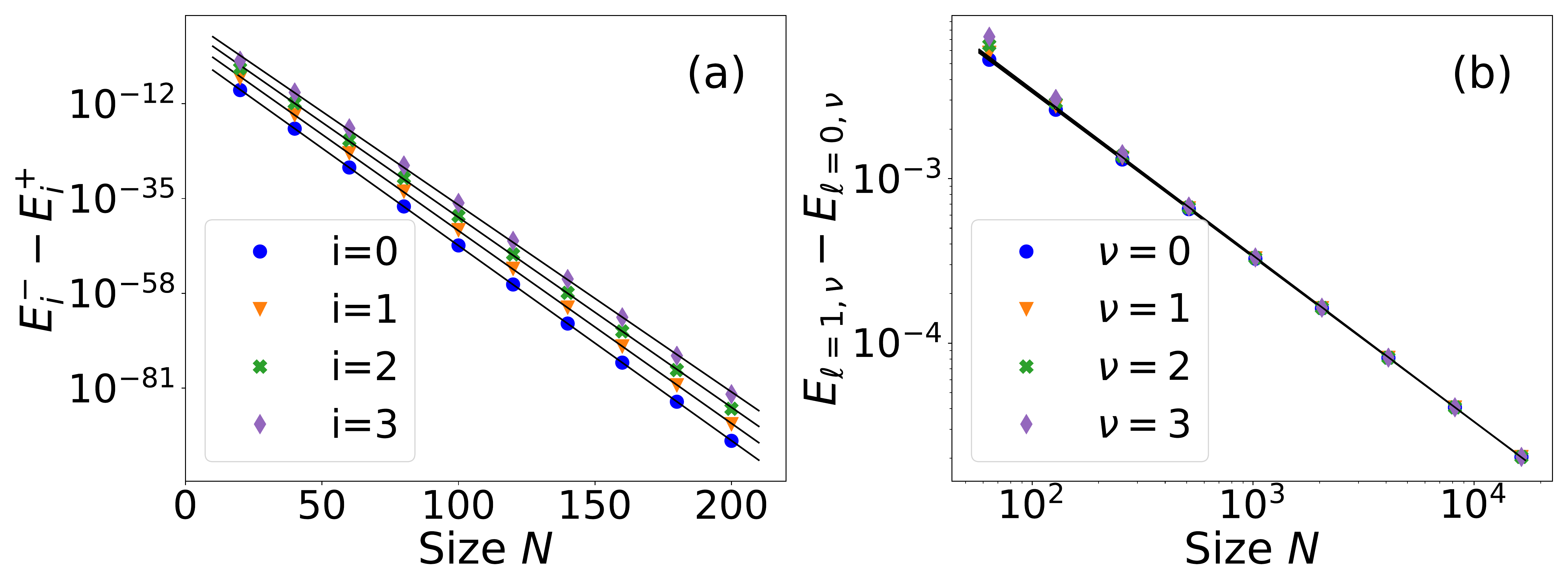}
\caption{\label{figdeltaE} Panel (a): Difference of energy  between even and odd energy levels in the LMG model, \(\Delta E_i = E_i^- - E_i^+\) with \(i = 0, 1, 2, 3\), as a function of the system size, \(N\), using lin-log axes. Black lines are the result of the fit of the depicted data to an exponential law. Panel (b): Difference of energy  between $\ell = 0$ and $\ell =1$ energy levels in the 2DVM model, \(\Delta E_\nu = E_\nu^{\ell = 1} - E_\nu^{\ell = 0}\) with \(\nu= 0, 1, 2, 3\) as a function of the system size, \(N\), using log-log axes. Black lines are the result of the fit of the depicted data to a power law. In both cases, the control parameter value is \(\xi = 0.5\).}
\end{figure*}

The difference between state degeneracy in the two models considered is clear from Fig.~\ref{figdeltaE}, where we plot the energy difference between the first four pairs of states with different symmetry for a control parameter $\xi= 0.5$ as a function of the system size The results for the LMG model are depicted in Fig.~\ref{figdeltaE}(a), computing the gap between even and odd states, while results for the 2DVM \(\ell = 0,1\) states are shown in  Fig.~\ref{figdeltaE}((b). The abscissa axis is linear in the LMG model and logarithmic in the 2DVM case. This implies an exponential decrease with system size of the energy splitting between even and odd states in the LMG model, while the corresponding energy difference in the 2DVM follows a power law with the system size. In the LMG case, we have used a library for real and complex floating-point arithmetic with arbitrary precision to achieve the required accuracy in the calculations~\cite{mpmath}.

\begin{figure*}
  \includegraphics[width=0.95\textwidth, angle=-0]{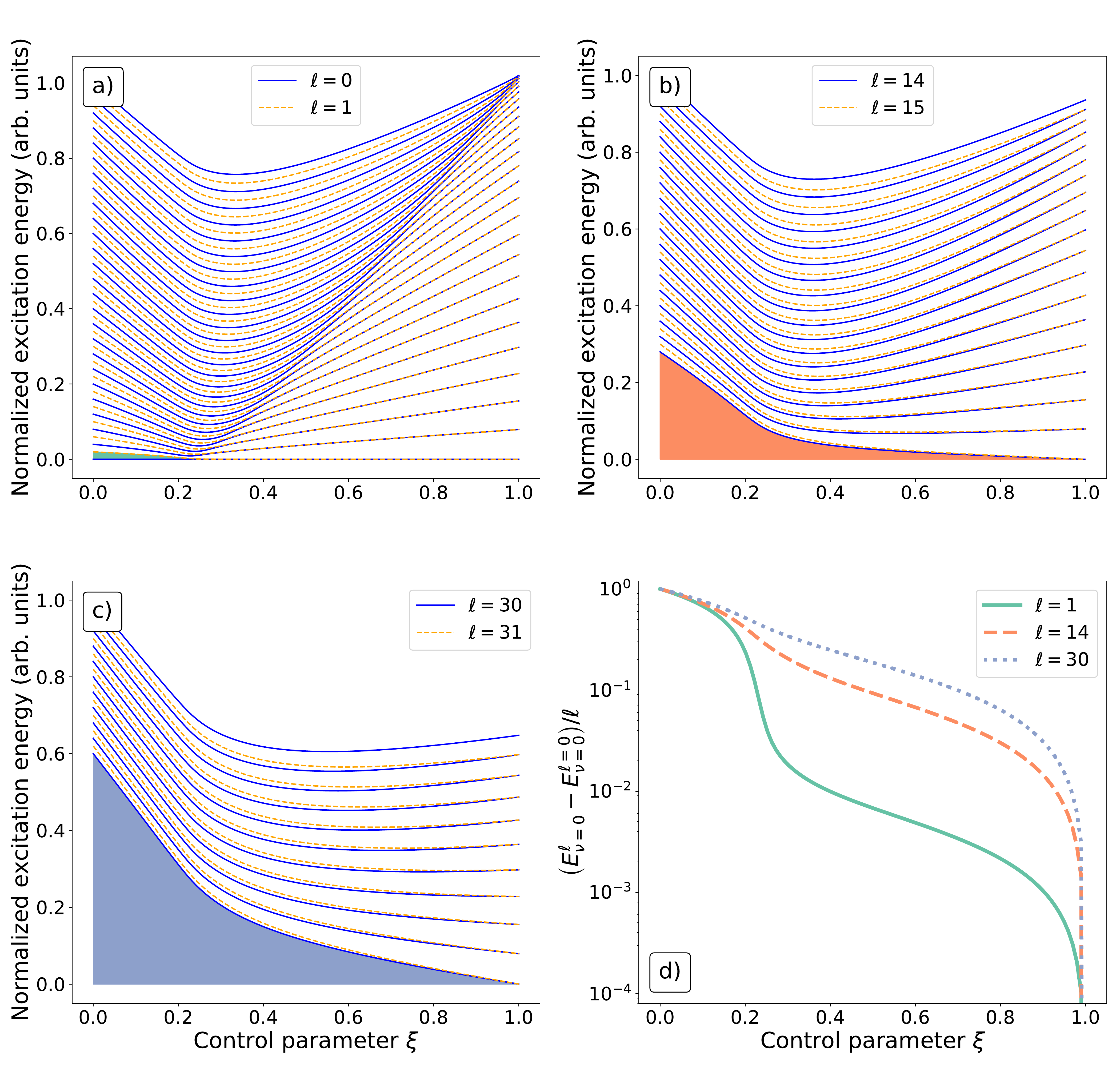}
\caption{\label{cent_barrier} Excitation energy values as a function of the control parameter \(\xi\) for the 2DVM Hamiltonian \eqref{2dvmi} with a system size \(N=50\) and angular momenta \(\ell = 0,1\) in panel (a), \(\ell = 14, 15\) in panel (b), and \(\ell = 30,31\) in panel (c). In panel (d), we show in log scale the excitation energy for minimum energy states of angular momenta $\ell=1$ (solid aqua line), $\ell=14$ (dash orange line), and $\ell=30$ (dotted violet line) with respect to the ground state, divided by the angular momentum value $\ell$. These differences have been highlighted in panels (a), (b), and (c) using the same color code.}
\end{figure*}

The  different results obtained for the two models can be understood considering the mean-field limit of their broken-symmetry phase. The LMG model can be mapped in the classical limit to an energy functional for zero momentum with three stationary points: two equivalent minima and a maximum between them that is located at the origin. Therefore, in one-dimensional models the tunneling between the two possible stationary solutions is exponentially suppressed with the barrier height. However, in higher dimensional models such as the 2DVM, the broken-symmetry phase  in the classical limit is mapped to a Mexican hat potential, with a maximum at the origin and a minimum with revolution symmetry at a given distance from the origin. In both cases, the critical energy of the ESQPT corresponds to the energy of the maximum at the origin and the energy functional associated with Hamiltonian operators \eqref{HLMGM} and \eqref{H2DVM} is a fourth-order function on the classical coordinate that depends on a single variable. However, for $n\ge 2$, the role of the angular momentum and the centrifugal barrier should be included into the picture. This is completely irrelevant for zero angular momentum states but, as the angular momentum value increases, the centrifugal barrier stymies the exploration of the maximum at the origin. This has already been noticed when studying the influence of angular momentum on other ESQPT precursors, as the participation ratio \cite{KRivera2022}. The effect of the centrifugal barrier can be clearly illustrated by plotting the correlation energy diagram for levels with high angular momenta, as shown in Fig.~\ref{cent_barrier} for the 2DVM with $N = 50$.  Levels in Fig.~\ref{cent_barrier}(a) have angular momentum \(\ell = 0,1\), in Fig.~\ref{cent_barrier}(b) the angular momentum is \(\ell = 14,15\), and in Fig.~\ref{cent_barrier}(c) \(\ell = 30,31\). It is clear how the level flow deviates from the $\ell = 0$ ESQPT separatrix for increasing  angular momentum values. In Fig.~\ref{cent_barrier}(d), the energy gap between the lowest energy state with bands $\ell=1$ (solid aqua line), $14$ (dash orange line), and $30$ (dotted violet line) and the ground state is depicted as a function of the $\xi$ control parameter value. These differences have been highlighted in panels (a), (b), and (c) using the same colors. For the sake of clarity, we  have scaled in Fig.~\ref{cent_barrier}(d) the energy difference by the $\ell$ angular momentum value, to make all curves start in unity and facilitate the comparison between the different case. Note that all the considered states are completely degenerate for $\xi=1$, in the dynamical symmetry limit.

\section{Application to a microcanonical out-of-time-order correlator}
\label{sec4}
The different nature of the degeneracy that occurs in the two models under consideration is expected to have a noticeable influence on the system dynamics. In order to illustrate this point, we consider the time evolution of a mOTOC. 

\begin{figure*}
\includegraphics[width=0.9\textwidth]{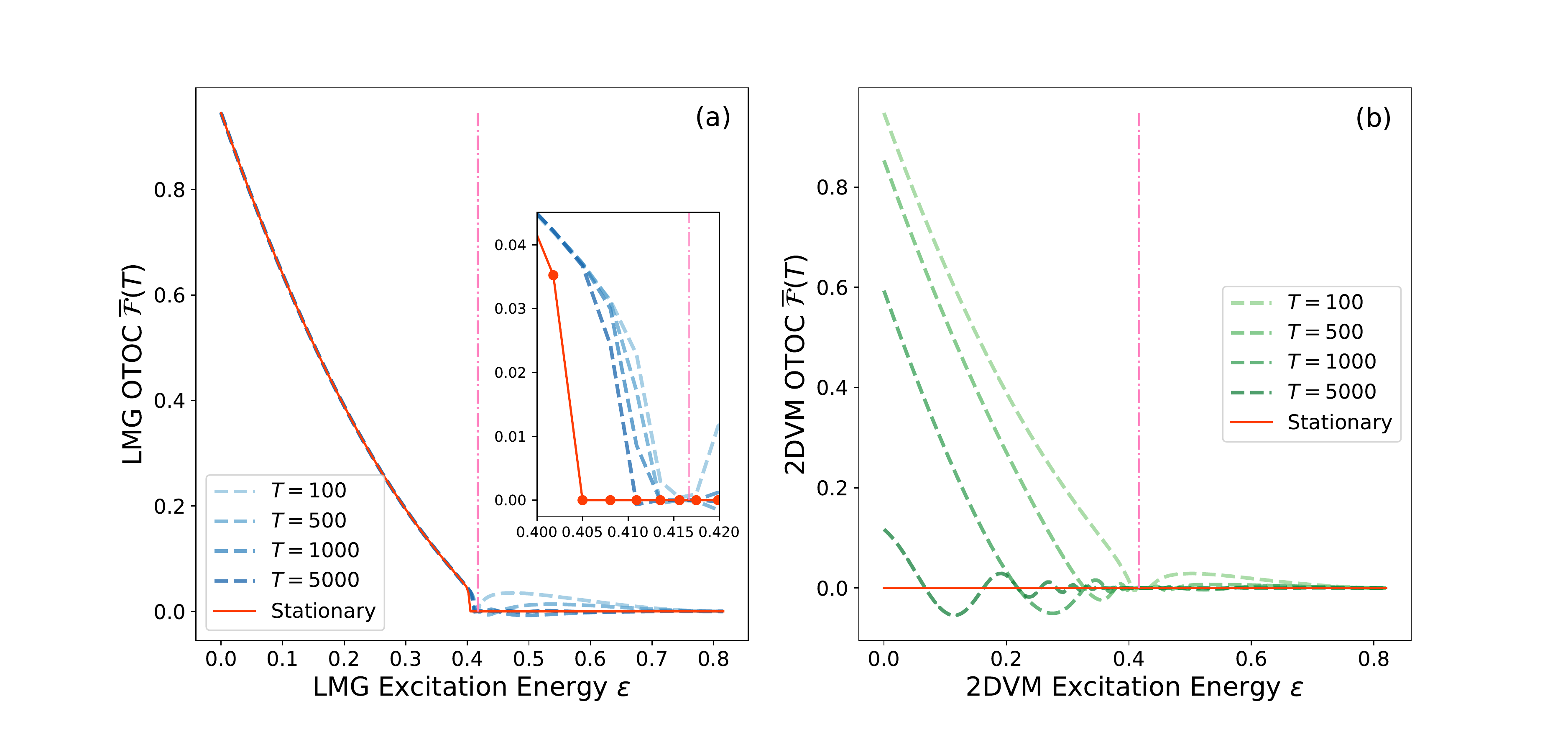}
\caption{\label{figFT} Panel (a): Time-averaged  mOTOC, \(\overline{{\cal F}^{(j)}_{VW}}(T)\), for the LMG model with \(\hat V =\hat W = \hat J_x\) as a function of the system's excitation energy scaled by the system size ($E/N$) for even parity eigenstates of Eq.~\eqref{HLMGM} Hamiltonian with a system size \(N = 300\). Panel (b): Time-averaged value  mOTOC, \(\overline{{\cal F}^{(j)}_{VW}}(T)\) for \(\ell = 0\) eigenstates of the 2DVM Hamiltonian in Eq.~\eqref{H2DVM} with \(N = 300\) with  \(\hat V = \hat D_-\) and \(\hat W = \hat D_+\). Results are plotted as a function of the system's excitation energy scaled by the system size ($(E-E_{gs})/N$). All panels:  The control parameter value is \(\xi = 0.6\) and the solid red line is the stationary value obtained with Eq.~\eqref{AvgF}. Dashed lines are the result of averaging for different time interval values  (see panel legend). The inset in the left panel is a zoom to the vicinity of the ESQPT critical energy. A vertical dot-dashed pink line marks the ESQPT critical energy in the mean-field limit.}
\end{figure*}

Out-of-time-order correlators (OTOCs) are four-point temporal correlation functions originally introduced in the context of superconductivity studies~\cite{Larkin1969}. In recent times, OTOCs have been in the limelight  for two key reasons \cite{Swingle2018}. On the first hand, they were suggested as an efficient quantum chaos probe once it was shown that OTOCs grow exponentially and in accordance with the system's Lyapunov exponent value at early times in non-integrable systems~\cite{Shenker2014, Maldacena2016, Hashimoto2017, Zhao2021}. In addition to this, it has also been found that OTOCs tend to their long-time limit saturation value in a characteristic way  in chaotic systems~\cite{IGarcia2018, Rammensee2018, Fortes2019, Novotny2023}. On the second hand, they can be used to quantify information scrambling in quantum systems, as they depend on the system entanglement spread \cite{Swingle2016a, Lewis2019, Xu2019, Niknam2020}. Despite the fact that the unusual time ordering of its constituents operators hinders the experimental access to OTOCs using local operators, several approaches using different platforms have successfully provided OTOC data \cite{Li2017, Garttner2017, Wei2018, Landsman2019, Pegahan2021, Green2022, Braumuller2022}. OTOCs have also been recently used to characterize QPTs \cite{Shen2017, Heyl2018, Ceren2019, Nie2020, Lewis2020, Bin2023}. Finally, it has been recently found that they  are also valuable ESQPT probes, because the instability associated with ESQPTs' critical points fosters an exponential OTOC increase at short times, even in integrable systems \cite{ChavezC2019, Pilatowsky2020, Xu2020, Hashimoto2020, Chavez2023}.

\begin{figure*}
\includegraphics[width=0.9\textwidth]{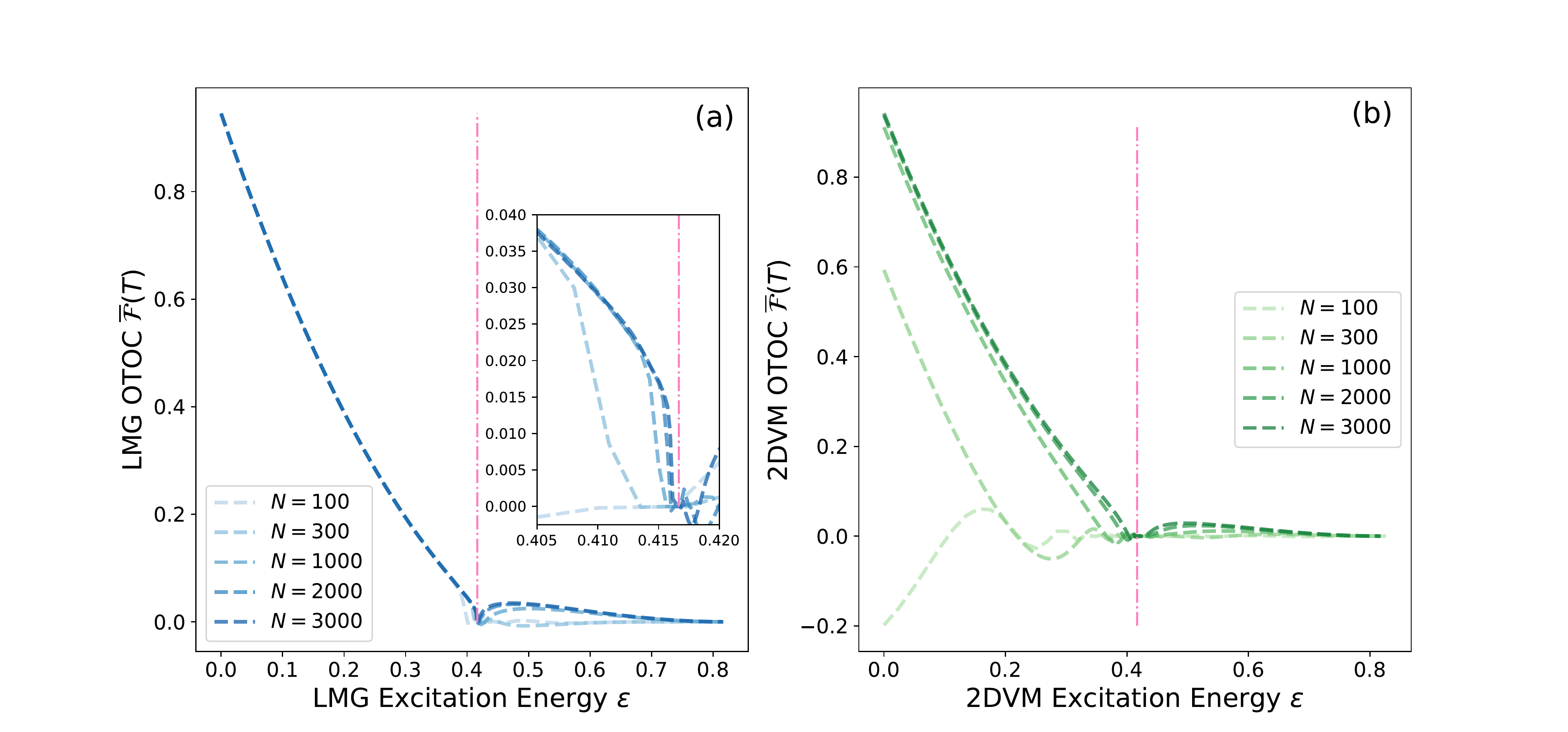}
\caption{\label{figFN}  Panel (a): Time-averaged value of the mOTOC, \(\overline{{\cal F}^{(j)}_{VW}}(T)\), for the LMG model with \(\hat V =\hat W = \hat J_x\) as a function of the system's excitation energy scaled by the system size ($(E-E_{gs})/N$) for even parity eigenstates of Eq.~\eqref{HLMGM} Hamiltonian for various system size values (see panel legend).  Panel (b):  Time-averaged value of the mOTOC, \(\overline{{\cal F}^{(j)}_{VW}}(T)\), for the 2DVM with  \(\hat V = \hat D_-\) and \(\hat W = \hat D_+\) as a function of the system's excitation energy scaled by the system size ($(E-E_{gs})/N$) for \(\ell = 0\) for eigenstates of the 2DVM Hamiltonian Eq.~\eqref{H2DVM} for various system size values, \(N\) (see panel legend).  
All panels: Calculations are carried out for a control parameter value \(\xi = 0.6\)  and the time average is performed over a time interval $T = 1000$. Dashed lines are the result for different system size values.  A vertical dot-dashed pink line marks the ESQPT critical energy in the mean-field limit.}
\end{figure*}

Given any two operators, $\hat{W}$ and $\hat{V}$, where $\hat{W}(t)$ is the operator $\hat{W}$ in the Heisenberg's representation, $\hat{W}(t)=e^{\imath\hat{H}t}\hat{W}e^{-\imath\hat{H}t}$, the spread of $\hat{W}(t)$ with $\hat{V}$ can be obtained through the expectation value of the squared commutator \cite{Swingle2018, Swingle2016a, Hashimoto2017}
\begin{equation}
{\cal C}_{w,v}(t)=\left\langle \left[\hat{W}(t),\hat{V}(0)\right]^{\dagger}\left[\hat{W}(t),\hat{V}(0)\right]\right\rangle~.
\label{C}
\end{equation}

The expectation value \eqref{C} is usually computed in the canonical ensemble. In the present case, we compute the microcanonical version of the correlator, the mOTOC, as the expectation value of the correlator over system eigenstates \cite{Hashimoto2017, Hashimoto2020}. The squared commutator Eq.~\eqref{C} can be expressed as ${\cal C}_{w,v}(t)= {\cal A}_{w,v}(t) -2\Re{[{\cal F}_{w,v}(t)]}$; the sum of a two-point correlator,  $ {\cal A}_{w,v}(t) = \left\langle \hat{W}^{\dagger}(t)\hat{V}^{\dagger}(0)\hat{V}(0)\hat{W}(t)\right\rangle + \left\langle \hat{V}^{\dagger}(0)\hat{W}^{\dagger}(t)\hat{W}(t)\hat{V}(0)\right\rangle$ and the real part of a four-point correlator, ${\cal F}_{w,v}(t)$, where out-of-time order effects take place
\begin{equation}
{\cal F}_{w,v}(t)=\left\langle \hat{W}^{\dagger}(t)\hat{V}^{\dagger}(0)\hat{W}(t)\hat{V}(0)\right\rangle.
\label{OTOC}
\end{equation}

Two of the authors (QW and FPB) have recently shown that the long-time average value of a mOTOC operator is a valid ESQPT order parameter for the LMG model Hamiltonian \eqref{HLMGM} \cite{Wang2019b}. The long-time average value of a similar OTOC has proved also a valid order parameter in an anharmonic version of the LMG model \cite{KRivera2022b}. In both cases, the average value of the mOTOC at the infinite time limit is able to distinguish between the dynamic phases below and above the ESQPT critical energy. In the present work, our aim is to show how different are the results obtained for the long-time mOTOC values depending on the number of effective degrees of freedom in the model, something that can be explained considering the different nature of the degeneracy below the ESQPT critical energy shown in Fig.~\ref{lmg_2dvm_ced}. For this purpose, we derive the formulas needed to compute the long-time average value of the mOTOC. Using Eq.~\eqref{OTOC} and the closure for the system eigenstates, \(|\psi_k\rangle\) with \(k = 1, \ldots, D\), the mOTOC for a system \(j\)-th eigenstate, \({\cal F}^{(j)}_{VW}(t)\), can be expressed as follows 
\begin{equation}
  {\cal F}^{(j)}_{VW}(t)  = 
\sum_{j_1,j_2,j_3=1}^D e^{i\omega(j, j_1,j_2,j_3)t}N^{(j)}_{j_1,j_2,j_3}~,\label{Fjt}
\end{equation}
where \(\omega(j, j_1,j_2,j_3) =  E_{j} + E_{j_2} - E_{j_1} - E_{j_3}\) and $N^{(j)}_{j_1,j_2,j_3}=\langle\psi_j|\hat W^\dagger|\psi_{j_1}\rangle\langle\psi_{j_1}|\hat V^\dagger|\psi_{j_2}\rangle\langle\psi_{j_2}|\hat W|\psi_{j_3}\rangle\langle\psi_{j_3}|\hat V|\psi_{j}\rangle$.

The time-averaged value of  \({\cal F}^{(j)}_{VW}(t)\) in the infinite time limit, denoted as \(\overline{{\cal F}^{(j)}_{VW}}\), is the equilibrium value of this four-point correlator
\begin{align}
  \overline{{\cal F}^{(j)}_{VW}}=&\lim_{T\to\infty}\frac{1}{T}\int_0^T {\cal F}^{(j)}_{VW}(t)dt=\lim_{T\to\infty}\overline{{\cal F}^{(j)}_{VW}}(T)\nonumber\\
  =& \sum_{j_1,j_2,j_3}^DN^{(j)}_{j_1,j_2,j_3}\delta_{\omega(j,j_1,j_2,j_3), 0}~.\label{AvgF}
\end{align}

Taking Eq.~\eqref{Fjt} into consideration, it is clear that a nonzero average value of \({\cal F}^{(j)}_{VW}(t)\) in the long-time limit implies that one or more instances of the  \(\omega(j,j_1,j_2,j_3)\) quantity should be equal to zero, to avoid oscillations. In order to check this, we have performed calculations for the LMG and 2DVM cases. Following Ref.~\cite{Wang2019b}, in the LMG model we define \(\hat V =\hat W = \hat J_x\). The use of Eq.~\eqref{AvgF} allows for a direct calculation of the long-time average value of the correlator, avoiding the evaluation of computationally expensive oscillatory time integrals. The results obtained  in the stationary \(t\to\infty\) limit  for the even parity states of the LMG Hamiltonian \eqref{HLMGM_boson} with a system size \(N = 300\) and a control parameter \(\xi = 0.6\) are shown in Fig.~\ref{figFT}(a) with a red solid line. This result agrees with the results obtained in  Ref.~\cite{Wang2019b}. If one computes intermediate average values for finite times, depicted with dashed blue lines in Fig.~\ref{figFT}(a), one can see how the results tend to the stationary solution as the averaging time increases. This is clearly seen in the inset included in this panel with a zoom to the results in the vicinity of the ESQPT critical energy. As already noticed in \cite{Wang2019b}, only eigenstates with energies less than the ESQPT critical energy, where even and odd states are degenerate, have a nonzero value of \(\overline{{\cal F}^{(j)}_{VW}}\), as shown in Fig.~\ref{lmg_2dvm_ced}(c). This can be explained taking into account that the operator \(\hat J_x\) transforms even parity states into odd parity ones (see Appendix ~\ref{app:atels}). Hence, from Eq.~\eqref{Fjt} for  \(\hat V =\hat W = \hat J_x\), it is clear that levels \(j\) and \(j_2\) are even parity states, while levels \(j_1\) and \(j_3\) are odd parity states. Therefore, the \(\omega(j,j_1,j_2,j_3) = 0\) condition can be fulfilled for degenerate even and odd parity eigenstates. It could also be zero whenever the sum of the energies of two even parity states is equal to the sum of the energies of two odd parity eigenstates, but this is something that does not occur for the LMG Hamiltonian \eqref{HLMGM}. The mean-field limit value of the critical energy is marked by a  dot-dashed pink vertical line in Fig.~\ref{figFT}(a).

 In the 2DVM case, an equivalent choice for the mOTOC operators would be \(\hat V = \hat D_-\) and \(\hat W = \hat D_+\). These two generators are part of the pairing operator and connect states with different values of the vibrational angular momentum,  \(\Delta\ell=\pm 1\) (see Appendix \ref{app:atels}). If this is the case, assuming that we compute the mOTOC for \(\ell = 0\) states, the fulfilment of the  \(\omega(j,j_1,j_2,j_3) = 0\) condition in Eq.~\eqref{AvgF} implies that the sum of energies of the \(j\)-th and \(j_2\)-th \(\ell = 0\) eigenstates should be equal to the sum of energies for the \(j_1\)-th and \(j_3\)-th \(|\ell| = 1\) eigenstates. From the results in panel (d) of Fig.~\ref{lmg_2dvm_ced}, it is clear that 2DVM  eigenstates belonging to Hamiltonian blocks with different values of the angular momentum are not degenerate for finite-size systems, though eigenstates with energies below the ESQPT critical energy can be close in energy. Due to this, no \(\omega(j,j_1,j_2,j_3)\) term is zero and the stationary value of the four-point correlator, \(\overline{{\cal F}^{(j)}_{VW}}\), is zero for all possible values of \(j\) (see solid red line in Fig.~\ref{figFT}(b)).  The small energy gaps occurring for eigenstates with energies under the ESQPT critical energy imply that it should take a longer time for them to reach the zero stationary limit value of the mOTOC, compared to what happens for states with energies above the ESQPT critical energy. This is clearly seen in Fig.~\ref{figFT}(b), where for shorter times (dashed lines with a lighter green shade) eigenstates above and below the critical energy can be distinguished. As in the previous case, the mean-field critical ESQPT energy is marked by a dot-dashed pink vertical line. As the time over which the average value is computed increases,  \(\overline{{\cal F}^{(j)}_{D_-D_+}}\to 0\) for all eigenstates. However, the small energy gap between levels with different angular momentum below the critical energy of the ESQPT indicates that \(\overline{{\cal F}^{(j)}_{D_-D_+}}\) could be considered as an approximate order parameter for the ESQPT, computing the average time value of \({\cal F}^{(j)}_{VW}(t)\) for finite time values and taking profit of the much longer time needed for states below the critical energy  to reach the mOTOC stationary value.

To check our previous results, we have performed a second set of calculations where we display \(\overline{{\cal F}^{(j)}_{VW}}(T)\)  for the two models under study, fixing the averaging time to a constant value, \(T= 1000\), and including results for different system sizes. The obtained results are shown in Fig.~\ref{figFN}. As in the previous figure, the results for the LMG model are shown in Fig.~\ref{figFN}(a), with different shades of blue for the different system size values, and the results for the 2DVM model are shown in Fig.~\ref{figFN}(b), using different shades of green. Again, as in Fig.~\ref{figFT}, we have marked the mean-field limit value of the critical energy with a vertical dot-dashed pink line. In the LMG case, there is no dependence on the system size for eigenstates with energies below the ESQPT critical energy, something that can be understood considering that these states are (up to the calculation numerical precision) degenerate for all system sizes considered. In Fig.~\ref{figFN}(a) inset  we show a zoom of the critical energy region where it can be clearly appreciated how, for larger system sizes, the time-averaged mOTOC goes to zero as one approach the mean-field critical value. Values of  \(\overline{{\cal F}^{(j)}_{VW}}(T)\) for eigenstates with energies larger than the ESQPT critical energy are closer to the expected stationary limit zero value for lower system sizes. This can  be understood considering that we are performing the averaging of the four-point mOTOC over a finite time value, \(T = 1000\), which is large enough to cancel the oscillatory integral for the smaller system sizes, but not for the larger ones. The results for the 2DVM case are shown in Fig.~\ref{figFN}(b). In this case, the smaller the system size, the closest to the zero stationary result for all eigenstates. Differences with the stationary value increase for increasing system sizes, in particular for states with energies less than the critical ESQPT energy. As in the LMG case the mOTOC at energies less than the critical energy tends to zero for increasing system sizes when the energy is close to the ESQPT critical energy. The difference with the zero stationary limit increases with system size for states with energies below the critical energy, due to the decreasing energy gap between states with different angular momentum values in this zone, as shown in Fig.~\ref{figdeltaE}(b). 

\section{Conclusions}
\label{Sec:Conclusions}

The present work shows that below the critical energy of an ESQPT, the nature of the degeneracy of the eigenvalues of two-level bosonic models depend on the number of degrees of freedom of the model. If the model has a single degree of freedom, neighboring levels belonging to different parity sectors approach each other exponentially as the system size increases, while the decrease of the energy splitting is only algebraically in models with  two or more degrees of freedom. Hence, in the latter case the pairs of states only become degenerate in the mean-field limit. This difference was illustrated with the LMG model and the 2DVM in the main text and with models with three and five collective degrees of freedom in the App.~\ref{appa}.

In the broken-symmetry phase of the 2DVM case, when considering all possible $\ell$ values, not all levels converge to the $\ell = 0$ state energy. This can be explained considering the influence of the centrifugal barrier. In fact, in models with two or more collective degrees of freedom ($n\ge 2$), states with angular momentum larger than zero only exactly converge to form degenerate rotational bands in the dynamical symmetry case (\(\xi = 1\)). For smaller values of \(\xi\), states with large angular momentum values do not follow the  $\ell = 0$  ESQPT separatrix line. 

The effects that the degeneracy dependence on the number of degrees of freedom have on the system dynamics were analyzed by computing the long-time average of a mOTOC. 
Our results make it clear that, as proposed in \cite{Wang2019b}, this quantity works as an order parameter for the  ESQPT in the LMG model 
even for {\em finite} system sizes. This finding can be extended to other one-dimensional models, where the energy separation between levels with different symmetries decreases exponentially as the system approaches the classical limit, as for example, the one-dimensional limit of the vibron model~\cite{bookmol,frank} or the one-dimensional bosonic pairing models \cite{Caprio2008}. Even in cases where the ESQPTs are associated with an infinite-dimension Hilbert space, the present results are expected to hold, such as the quantum quartic oscillator~\cite{Cejnar2008} or the squeeze-driven Kerr oscillator \cite{Chavez2023}, where the exponential decrease of the energy difference has been shown experimentally~\cite{Frattiniarxiv}. However, in models with two or more collective degrees of freedom, the mOTOC can only be considered an order parameter in the mean-field limit, and only for states with angular momentum values low enough to converge on the corresponding \(\ell = 0\) rotational band head.

\begin{acknowledgments}
We wish to acknowledge useful discussions with José Miguel Arias, Pedro
  Pérez Fernández, Jorge Dukelsky, Armando Relaño, Ángel López Corps, Jorge Hirsch, and Saúl Pilatowski-Cameo.

  This project has received funding from the Grant PID2022-136228NB-C21 funded by MICIU/AEI/
10.13039/501100011033 and, as appropriate, by “ERDF A way of making Europe,” by
“ERDF/EU,” by the “European Union,” or by the “European Union
NextGenerationEU/PRTR.” This work
  has also been partially supported by the Consejer\'{\i}a de
  Conocimiento, Investigaci\'on y Universidad, Junta de Andaluc\'{\i}a
  and European Regional Development Fund (ERDF), UHU-1262561 (JKR and
  FPB) and US-1380840 (JKR), and PY2000764. JKR also acknowledges support from a Spanish Ministerio de Universidades ``Margarita Salas'' Fellowship. Computing resources supporting this work were provided by the CEAFMC and Universidad de Huelva High
  Performance Computer (HPC@UHU) located in the Campus Universitario ``El Carmen'' and funded by FEDER/MINECO project UNHU-15CE-2848. This research was also supported by the
NSF CCI grant (Award Number 2124511). 
\end{acknowledgments}

\appendix

\section{Extension to models in \(n = 3\) and \(n = 5\) dimensions}
\label{appa}
  The vibron model (VM) was introduced in the 80s by Iachello, extending the algebraic approach to the study of molecular structure \cite{frank,bookmol}. In particular, ro-vibrational excitations for a diatomic molecule are treated as bosonic collective excitations \cite{bookmol}. Due to the three-dimensional nature of the problem associated with the dipole degree of freedom in diatomic molecules, this case has \(n=3\) and a \(u(4)\) Lie algebra as its dynamical algebra.  The bosonic operators requested to build the sixteen generators of the dynamical algebra  are a scalar boson operator \(s^\dagger (s)\) and an angular momentum one boson \(p_\mu^\dagger (p_\mu)\) with \(\mu = \pm 1, 0\). As in the previous cases, the \(u(4)\) generators are built as bilinear products of creation and annihilation operators \cite{frank, bookmol}.  As in the two previous cases, the VM has two dynamical symmetries 
   converging in $so(3)$, the system's symmetry algebra, associated with the conservation of the angular momentum
\begin{align}
&u(4)\supset u(3) \supset so(3)&\mbox{Chain (VM-I)}~,\label{vmi}\\
&u(4)\supset so(4)\supset so(3) &\mbox{Chain (VM-II)}~.\label{vmii}
\end{align}
\noindent In this case, the total number of \(s\) and \(p\) bosons is denoted as \(N\) and $[N]$ corresponds to the totally symmetric \(u(4)\) irrep that spans the system's Hilbert space. The dynamical symmetry in Eq.~\eqref{vmi} is a convenient approximation to model vibration of floppy, weakly-bent molecules and the Eq.~\eqref{vmii} dynamical symmetry provides a Morse-like spectrum and it has been applied to many molecular species \cite{bookmol}. A model Hamiltonian defined in the same way as in the previous two cases can be built using the first order Casimir operator of the \(u(3)\) subalgebra in Eq.~\eqref{vmi} and the \(so(4)\) pairing operator, built with the second order Casimir operator of \(so(4)\) in Eq.~\eqref{vmii}

\begin{equation}
  \hat H_{VM} = (1-\xi) \hat n_p + \frac{\xi}{N}  \hat P_p~,
  \label{HVM}
\end{equation}
\noindent with a control parameter \(\xi\in[0,1]\), \(\hat n_\tau = \sum_\mu p^\dagger_\mu p_\mu \), and \(\hat P_p =  N(N+2) - \hat D^2 - \hat J^2\) where \(\hat D^2 + \hat J^2\) is the second order Casimir operator of the \(so(4)\) subalgebra \cite{frank, Iachello1996,bookmol, bookalg}. In this case the symmetry algebra is \(so(3)\), due to conservation of angular momentum,  and Hamiltonian \eqref{HVM} is split into different blocks, one for each angular momentum value, \(J\), considered \cite{bookmol}.

\begin{figure*}
  \includegraphics[width=0.75\textwidth, angle=-0]{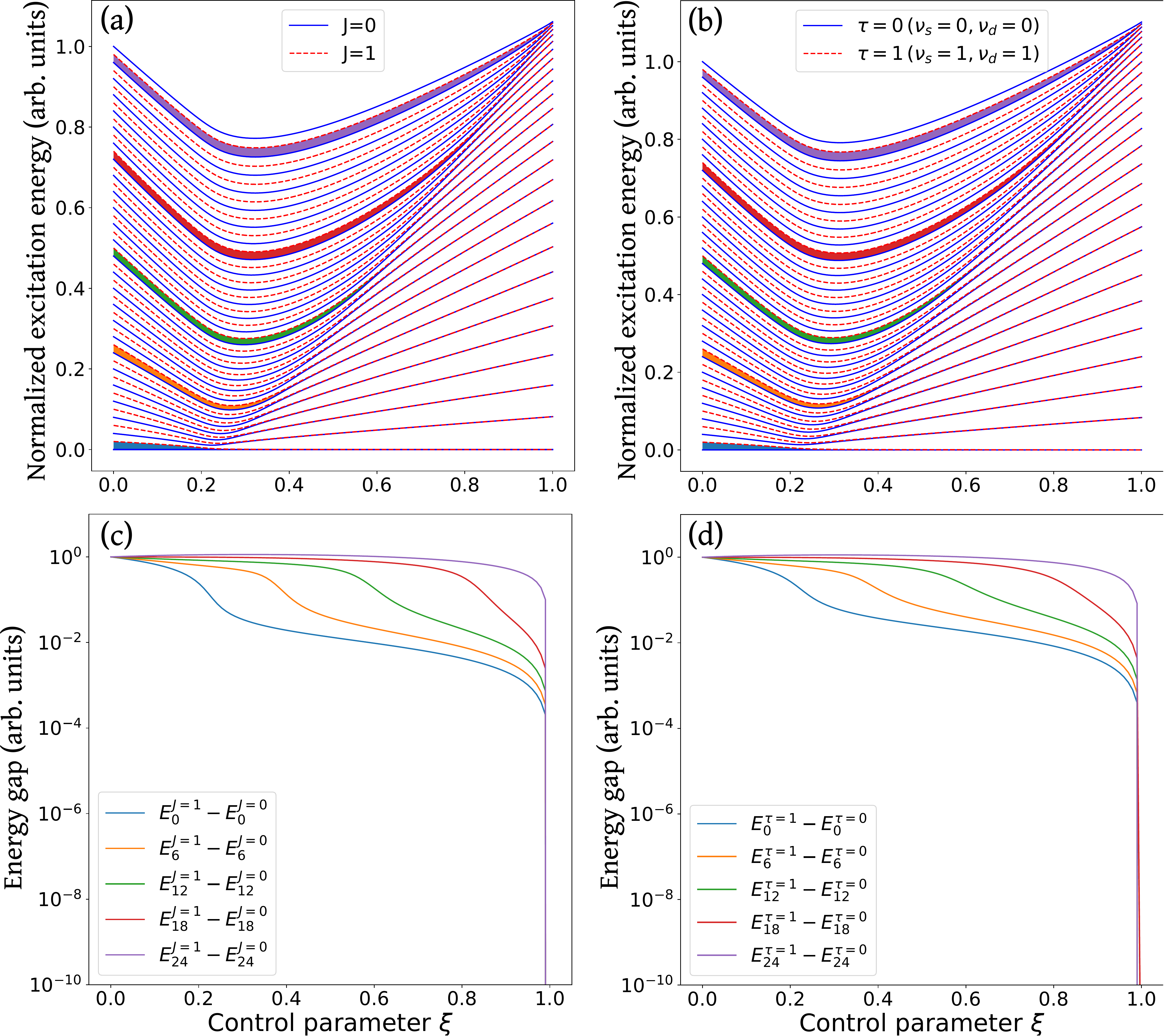}
\caption{\label{VM-IBM_spectra} Panel (a): Excitation energy scaled by the system size ($\varepsilon = (E-E_0)/N$) as a function of the control parameter $\xi$ for the VM with model Hamiltonian Eq.~\eqref{HVM} and system size \(N = 50\). Solid blue (dashed red) lines mark levels with angular momentum $J=0 (1)$.  Panel (b): Excitation energy scaled by the system size ($\varepsilon = (E-E_0)/N$) as a function of the control parameter $\xi$ for the IBM model Hamiltonian Eq.~\eqref{HIBM} with a system size \(N = 50\). Solid blue (dashed red) lines mark levels with seniority \(\nu_s=\nu_d=0 (\nu_s=\nu_d=1)\). In both panels the color-filled area marks the energy difference  between selected states with different angular momentum (panel (a)) or seniority (panel (b)). Panel (c): Energy difference between selected pairs of states of the VM Hamiltonian Eq.~\eqref{HVM} having different angular momentum as a function of the control parameter $\xi$. Panel (d): Energy difference between selected states of Hamiltonian Eq.~\eqref{HIBM} with different seniority as a function of the control parameter $\xi$. In both cases, the labels of the selected pairs of levels are provided in the legend of the panels and the color used for each pair of states is the same color used to fill the corresponding area in the upper panel.}
\end{figure*}

The model Hamiltonian \eqref{HVM} has a second-order ground-state QPT between the \(u(3)\) and \(so(4)\) dynamical symmetries, and an associated ESQPT.  The correlation energy diagram is displayed in the panel (a) of Fig.~\ref{VM-IBM_spectra} for levels with angular momentum $J=0$ (solid blue lines) and $J=1$ (dashed red lines). The correlation energy diagram is akin to the correlation energy diagrams of the LMG and the 2DVM shown in Fig.~\ref{lmg_2dvm_ced}. Once again, the difference between levels with different angular momentum is highlighted. In the panel (c) of Fig.~\ref{VM-IBM_spectra} these differences are plotted using semi-log scale. As we observed in the 2DVM, the order of the degeneracy does not reach the numerical precision until $\xi=1$.

The IBM has a very successful history in the study of nuclear structure using algebraic methods. This  model was introduced in the 70's of the XX-th century by Arima and Iachello  \cite{[][{, and references therein.}]booknuc}. Since then, it has become a standard tool in the study of nuclear structure from a collective point of view \cite{booknuc}. In this case, \(n = 5\) due to the  five dimensions that characterize the nuclear collective problem   and the model dynamical algebra is \(u(6)\)  \cite{booknuc}. The nuclear excitations are treated as bosons that can be traced back to coupled nucleons with angular momentum zero or two. Therefore the model building blocks are a scalar boson, \(s\) and a boson with angular momentum two, \(d_\mu\), with \(\mu = \pm 2, \pm 1, 0\) and the 36 generators of \(u(6)\) are expressed as the bilinear products of creation and annihilation boson operators \cite{booknuc,frank,bookalg}. As in the previous case, the angular momentum is conserved in this case and, therefore, the system's symmetry algebra is \(so(3)\). In this case there are at least three relevant dynamical symmetries, but we concentrate in this work on the two subalgebra chains that are equivalent to the cases previously mentioned
\begin{align}
&u(6)\supset u(5) \supset so(5)\supset so(3)&\mbox{Chain (IBM-I)}~,\label{ibi}\\
&u(6)\supset so(6)\supset so(5)\supset so(3) &\mbox{Chain (IBM-II)}~.\label{ibii}
\end{align}
\noindent In this case, the total number of \(s\) and \(d\) bosons is denoted as \(N\) and it defines the totally symmetric \(u(6)\) irrep that spans the system's Hilbert space. The dynamical symmetry in Eq.~\eqref{ibi} is a convenient approximation to model nuclear structure in spherically symmetric (vibrational) nuclides, while the dynamical symmetry Eq.~\eqref{ibii} provides a way to model the so called gamma-unstable nuclei \cite{frank, booknuc}. Hence, we define a model Hamiltonian in the same way as in the previous cases.  The first term, \(\hat n_d\), is the number operator of \(d\) bosons, which is the first order Casimir operator of \(u(5)\)  in Eq.~\eqref{ibi}. The second term is the Casimir operator of the $so(6)$ subalgebra in Eq.~\eqref{ibii}, \(\hat P_d = 2\left(N(N+4) - \left[ d^\dagger\cdot d^\dagger- s^\dagger s^\dagger \right]\left[\tilde d\cdot\tilde d-\tilde s\tilde s\right]\right)\),
\begin{equation}
  \hat H_{IBM} = (1-\xi) \hat n_d + \frac{\xi}{N}  \hat P_d~.
  \label{HIBM}
\end{equation}
As in the three previous cases, the Hamiltonian has one control parameter, \(\xi\in[0,1]\) \cite{frank, booknuc, bookalg}. If we do not consider other possible dynamical symmetries, the symmetry algebra is \(so(5)\) instead of \(so(3)\), and the conserved quantity is the seniority ($\nu_s$ and $\nu_d$), which is related with the label \(\tau\) of the irrep. This implies that Hamiltonian Eq.~\eqref{HIBM} can be split into seniority blocks \cite{frank, booknuc, bookalg}. As in the previous cases, the model Hamiltonian \eqref{HIBM} has a second-order ground-state QPT between the \(u(5)\) and \(so(6)\) dynamical symmetries, and the corresponding ESQPT.  The correlation energy diagram is plotted in panel (b) of Fig.~\ref{VM-IBM_spectra}, for \(\tau = 0\) and \(1\)  levels. On more time, a very similar spectra is obtained, where the eigenvalues seem to be degenerate in the broken-symmetry phase. However, the highlighted differences in the panel (b) are plotted using the log-linear scale in panel (d) of Fig.~\ref{VM-IBM_spectra}. As it was expected, the degeneracy  is not achieved until the system is in the dynamical symmetry $so(6)$.

\section{Matrix elements of the relevant operators of the $u(2)$ and $u(3)$ models}
\label{app:atels}
In this section we provide of the matrix elements that are needed to develop the calculations presented in the present manuscript.
\subsection{Matrix elements of the LMG model}
The chain I of the LMG model, introduced in Eq.~\eqref{lmgmi}, provides one label $n_t$ to name the states of the basis $\left\{\ket{[N]n_t},\;n_t=0,1...,N\right\}$. The elements of this basis conserve the parity, $\hat{\Pi}\ket{[N]n_t}=(-1)^{n_t\%2}\ket{[N]n_t}$, where the symbol $\%$ denotes the modulo-$2$ operation. The expected value of the operator $\hat{J}_x$ in this basis is
\begin{align}\label{Jxmatel}
\braket{[N]n_t'}{\hat{J}_x}{[N]n_t}=&\frac{1}{2}\sqrt{(N-n_t)(n_t+1)}\delta_{n_t',n_t+1} \nonumber \\
+&\frac{1}{2}\sqrt{(N-n_t-1)n_t}\delta_{n_t',n_t-1} ~.
\end{align}
 From Eq.~\eqref{Jxmatel} it is trivial to realize that operator $\hat{J}_x$ mixes elements with different parity, however $\hat{J}_x^2$ connects the state $\ket{[N]n_t}$ with itself and with $\ket{[N]n_t\pm 2}$, both with the same parity.

\subsection{Matrix element of the 2DVM}
The relevant operators in this model are $\hat{n}$, $\hat{D}_{\pm}$, $\hat{\ell}$, and $\hat{W}^2$, the latter can be expressed as $\hat{W}^2=\frac{1}{2}\left(\hat{D}_+\hat{D}_-+\hat{D}_-\hat{D}_+\right)+\hat{\ell}^2$. The basis most frequently used is the one associated with chain I, Eq.~\eqref{2dvmi}. The element of this basis can be labeled using the vibrational quantum number $n$ and the vibrational angular momentum $\ell$ as $\left\{\ket{[N]n,\ell}\equiv \ket{n^{\ell}}\right\}$ with $n=N, N-1, N-2, ..., 0$ and  $\ell = \pm n, \pm(n-2), ..., \pm (n\mod 2)$. The matrix elements of the operators $\hat{n}$ and $\hat{\ell}$ are trivial, and the matrix elements of $\hat{D}_{\pm}$ are
\begin{align}\label{Dpmmatel}
    \braket{n'^{\ell'}}{\hat{D}_\pm}{n^{\ell}}=\pm\sqrt{(N-n)(n\pm\ell+2)} \delta_{\ell',\ell\pm1}\delta_{n',n+1}~.
\end{align}
From Eq.~\eqref{Dpmmatel} it is  understood how $\hat{D}_{\pm}$ connect a state with angular momentum $\ell$ to another with $\ell'=\ell\pm 1$. The matrix elements of $\hat{W}^2$ can be easily derived from Eq.~\eqref{Dpmmatel} and preserve the vibrational angular momentum $\ell$.

\bibliography{letter_OTOC}

\end{document}